%% file: main.tex
 \documentclass[acmlarge]{acmart}

\input{macros}

\makeatletter
\newcommand{\myconfshort}{\acmConference@shortname}
\newcommand{\myconffull}{\acmConference@name}
\newcommand{\myconfdate}{\acmConference@date}
\newcommand{\myconfloc}{\acmConference@venue}
\AtBeginDocument{
  \fancypagestyle{firstpagestyle}{
    \fancyhead{}%
    \fancyfoot[C]{}%
  }
  \fancyhf{}
  \fancyhead[LO]{\@headfootfont\shorttitle}%
  \fancyhead[RE]{\@headfootfont\@shortauthors}%
  \fancyhead[LE]{\@headfootfont\footnotesize \myconfshort, \myconfdate, \myconfloc}%
  \fancyhead[RO]{\@headfootfont\footnotesize \myconfshort, \myconfdate, \myconfloc}%
  \fancyfoot[C]{}%
}
\makeatother
\acmBooktitle{\conffull\@ (\confshort), \confdate, \confloc}

\copyrightyear{2026}
\acmYear{2026}
\setcopyright{cc}
\setcctype{by}
\acmConference[FAccT '26]{The 2026 ACM Conference on Fairness, Accountability, and Transparency}{June 25--28, 2026}{Montreal, QC, Canada}
\acmBooktitle{The 2026 ACM Conference on Fairness, Accountability, and Transparency (FAccT '26), June 25--28, 2026, Montreal, QC, Canada}
\acmDOI{10.1145/3805689.3806724}
\acmISBN{979-8-4007-2596-8/2026/06}

\begin{document}

\title{What Influences Readers' and Writers' Perceived Necessity of AI Disclosure?}


\author{Jingchao Fang}
\email{jcfang@uchicago.edu}
\orcid{0000-0002-9412-4244}
\affiliation{
  \institution{University of Chicago}
  \city{Chicago}
  \state{Illinois}
  \country{United States}}

\author{Victoria Xiaohan Wen}
\email{wen7@uchicago.edu}
\orcid{0009-0009-5249-5045}
\affiliation{
  \institution{University of Chicago}
  \city{Chicago}
  \state{Illinois}
  \country{United States}}

\author{Mina Lee}
\email{mnlee@uchicago.edu}
\orcid{0000-0002-0428-4720}
\affiliation{
  \institution{University of Chicago}
  \city{Chicago}
  \state{Illinois}
  \country{United States}}

\renewcommand{\shortauthors}{Fang et al.}

\input{abstract}




\maketitle

\input{introduction}
\input{background}

\input{factors_hypo}

\input{method}

\input{results}

\input{discussion}

\input{conclusion}
\input{endmatter}


\bibliographystyle{ACM-Reference-Format}
\bibliography{main}

\input{appendix}

\end{document}

%% file: macros.tex
\usepackage[normalem]{ulem} 

\usepackage{multirow}
\usepackage{booktabs}
\usepackage{tikz}
\usepackage{xcolor}
\usepackage{longtable}
\usepackage{array}
\usepackage{makecell}
\usepackage[dvipsnames]{xcolor}
\usepackage{subcaption}
\newcommand{\smalladdlinespace}{\vspace{4pt}}
\usepackage{colortbl}
\usepackage{diagbox}

\definecolor{lightpink}{RGB}{255, 224, 222}

\definecolor{lightblue}{RGB}{212, 233, 255}

\definecolor{darkblue}{RGB}{39, 125, 217}

\definecolor{lightpurple}{RGB}{179, 55, 142}

\definecolor{darkpurple}{RGB}{102, 38, 181}

\definecolor{darkgreen}{RGB}{49, 130, 53}

\definecolor{amethyst}{RGB}{138, 88, 245}

\newcommand{\CR}[1]{{\color{black} #1}}

%% file: abstract.tex
\begin{abstract}


The growing capability of artificial intelligence (AI) leads to its increasing adoption in writing, spurring discussions around whether writers should disclose their AI use in writing. What influences the perceived necessity of disclosure? We look into this question from three dimensions: perspective (reader or writer of the text), purpose (the goal of reading or writing), and procedural factors (how AI was used in the writing process in terms of replaceability, effortfulness, intentionality, and directness). In a vignette study ($N=727$), we find that readers consider disclosure to be more necessary than writers, and disclosure is regarded as more necessary when AI's contribution in writing is irreplaceable, directly incorporated, and when the writer does not intentionally steer AI generation. To our surprise, the writers' intentionality of AI use produces contrasting effects on readers' and writers' perceived necessity of disclosure. 
Moreover, the effort of writing shows no significant effect on the perceived necessity.
This study contributes to the conversation on transparent AI use by revealing readers' and writers' grassroots judgments, providing a unique angle to reflect on existing regulations, and offering insights into how AI disclosure guidance and tools could be designed to better align with readers' and writers' perceptions.

\end{abstract}

%% file: introduction.tex
\section{Introduction}

Many people are turning to AI to assist with their writing process~\cite{yang2022ai,
liang2024mapping, calderwood2020novelists}, with goals ranging from grammar
check to automated content generation, as well as practices that span using AI-generated
text as a reference to adopting it verbatim~\cite{shi2022effidit, lee2024design,
xu2025patterns}.
Today, AI-generated text is often indistinguishable from human-written text,
especially when AI-generated text is partially and incrementally blended into a
larger piece of writing~\cite{zhang2024llm, gao2023comparing}. While this marks an
exciting advancement of AI technologies, it makes it extremely challenging for
readers to tell who created the content that they are reading~\cite{jakesch2023human}.
Blurred creatorship makes it difficult for readers to contextualize the text,
discern its intent, and undermine readers' ability to judge its accountability
and trustworthiness of writing content~\cite{toff2024or}. Furthermore, numerous
studies show that AI may hallucinate (i.e., generate false information that sounds
plausible) and undisclosed AI-generated text may be weaponized to exacerbate the
spread of misinformation~\cite{ji2023survey, zhang2025siren, pan2023risk}.
Given that AI generation detectors fall short in accuracy and reliability,
especially as AI models and writers' practices (e.g., prompting and editing
strategies) continue to evolve~\cite{kushnareva2023ai, zhang2024llm,
krishna2023paraphrasing}, transparent AI use relies largely on writers' \textbf{AI
disclosure}---the practice of disclosing AI contribution in content creation~\cite{renieris2024artificial,
weaver2024artificial}.

Disclosing sources of contributions is well established in human writing; clearly
acknowledging human contributors in a collaborative work has long been an ethical
and moral norm~\cite{scratchmanualcredit, authorshipcontributiondisclosure,
icmjeauthorshipcontributors}.
However, disclosing AI use in writing has yet to become a common practice~\cite{draxler2024ai}.
As we are still in the early phase of adapting to an ecosystem with a
significant presence of AI involvement, only a small portion of platforms,
publishers, and organizations have started to explicitly specify the acceptable use
cases of AI and standards of AI disclosure (e.g.,~\cite{acmpolicy, springernaturepolicy}).
In most cases, however, the expectation of AI disclosure remains ambiguous. Under
such circumstances, some writers choose not to disclose their AI use, or even intentionally
hide it, due to concerns such as their work being devalued~\cite{zhang2025secret,
adnin2025eduperspective}.
This writers' tendency toward non-disclosure diverges from readers' expectations.
For instance, most news readers want journalists to disclose how AI is used in newsrooms~\cite{newresearch}.
While there appears to be a clear gap between readers' expectations and writers'
practices, it is not yet clear whether there is an underlying gap between readers'
and writers' perceptions of the necessity of AI disclosure.


With the ultimate goal of aligning disclosure behaviors with ideal practices
, we need to first understand the expectations of AI disclosure from the
standpoints of readers and writers.
Currently, AI disclosure guidelines (which represent the ``\textit{necessity}''
of AI disclosure) are usually formulated through a top-down approach by people who
implement and enforce regulations (e.g., policymakers, publishers, or community
leaders and moderators of content-sharing platforms), who are not the ones
directly affected by the writings and AI disclosure~\cite{morosoli2025public}. In this study, we take a bottom-up approach by exploring readers' and writers' ``\textit{perceived necessity}'' of AI disclosure, revealing the grassroots perceptions of
stakeholders who directly interact with the writings and are affected by AI
disclosure.
We believe that our readers- and writers-centered inquiry into AI use transparency is valuable for the FAccT community as it enriches the discussion
around AI transparency. As~\citet{corbett2023interrogating} pointed out, transparency research in FAccT (1) has mostly focused on algorithmic transparency provided by
regulators or technical experts, overlooking AI users' approach to transparent AI
use (e.g., writers in our study context); (2) should place greater emphasis on
bridging the gulf between providers of transparency (e.g., writers) and
recipients of transparency (e.g., readers); (3) should provide transparency in a
way grounded in the needs of end-users (e.g., readers and writers, as opposed to
regulators or technical experts). 
\CR{While some recent work is related to (1) and (3)---for example, \citet{he2025deservecredit} designed an attribution toolkit and \citet{hoque2024hallmark} proposed an interactive system to facilitate writers’ disclosure---these efforts focus on the providers of transparency, leaving the gap between transparency providers (writers) and transparency recipients (readers) underexplored. We take a step toward exploring this missed opportunity in the space of AI disclosure.}

Based on prior works in transparency and disclosure in AI-assisted writing~\cite{fang2025shapes,
he2025deservecredit, khosrowi2023diffusing, zhang2025secret}, we synthesize three
dimensions that may influence the perceptions around AI disclosure: the \emph{perspective} (i.e., reader or writer of the text),
the \emph{purpose of the written text} (i.e., the goal of reading or writing the
text), and \emph{procedural factors} (i.e., how AI was used). Concretely,
procedural factors include \textit{replaceability} (the extent to which AI's contribution can be replaced), \textit{effortfulness} (how much effort the writer put into writing), \textit{intentionality} (how intentionally the writer steered AI generation), and \textit{directness} (how much AI-generated content is directly incorporated in the final writing), capturing varied AI involvement in writing. 
Our research questions are:
how do the \emph{perspective} \textbf{(RQ1)}, \emph{purpose of the written text} \textbf{(RQ2)}, and
\emph{procedural factors} \textbf{(RQ3)} affect the perceived necessity of AI disclosure?
We conduct a vignette study with 727 participants to investigate these questions. Concretely, we operationalize the combinations of perspective, purpose, and
procedural factors into vignettes representing a wide range of hypothetical
reading and writing situations,  
and ask study participants whether they perceive AI disclosure as necessary in
each situation.




We find that perspective and procedural factors individually and collectively
shaped readers' and writers' perceived necessity of AI disclosure.  
Readers were more likely to think disclosure is necessary than
writers (\textit{perspective}), while the purpose of the written text did not
have a main effect. 
Disclosure was more likely to be perceived as necessary when AI involvement is
significant. Specifically, a higher necessity
of disclosure was perceived when AI contribution is irreplaceable by a writer's own work (\textit{procedural
factors--low replaceability}), when a writer lets AI take the lead in writing (\textit{procedural
factors--low intentionality}), and when a large amount of AI-generated content is
directly incorporated into the final writing outcome (\textit{procedural factors--high directness}). Surprisingly, the perceived necessity of disclosure did not increase significantly when a writer puts little effort into writing (\textit{procedural
factors--low effortfulness}), countering a commonly held assumption that less effort leads to less ownership and necessitates AI disclosure.
The effect of procedural factors on perceived necessity was moderated
by perspectives. The writers' intentionality of AI use shaped readers' and
writers' perceived necessity of disclosure in opposing ways: low intentionality increased readers' perceived necessity of AI disclosure but decreased writers' perceived necessity of AI disclosure. 
This prompts us to reflect on the need
for motivating writers' AI disclosure when it is desired by readers. We conclude by discussing two future directions for motivating AI disclosure: scaffolding writers' AI disclosure and designing behavioral interventions that emphasize factors that increase writers' perceived necessity of disclosure.

%% file: background.tex
\section{Background}

\subsection{Human-AI collaborative creation \CR{and AI-assisted writing}}\label{sec:background_cocreation}


The rise of AI capabilities and applications enables human-AI collaborative creation (co-creation) of texts~\cite{lee2024design, gero2022sparks, hwang202580}, music~\cite{fu2025collaborativemusiccocreation, micchi2021keep, newman2023human}, images~\cite{fan2024contextcam, oppenlaender2025artworks, lawton2023tool}, and videos~\cite{huh2025videodiff, choi2024vivid, liu2023ai}. 
AI-assisted writing is a type of content co-creation that becomes increasingly common across academic~\cite{amirjalili2024exploring, adnin2025eduperspective}, professional~\cite{liang2024mapping, liang2024monitoring, long2023tweetorial, liang2025widespreadadoptionlargelanguage, hanley2023machine}, and personal contexts~\cite{liu2024artificial, hohenstein2023artificial, sun2024we}. 
\CR{Literature in writing theory commonly views writing as a process involving recursive cognitive activity rather than merely a product or an output~\cite{murray1972teach, hairston1982winds}.
Today, writers can use AI to} support in all stages of their writing process~\cite{flower1981cognitive}, from planning and ideation~\cite{gerothesaurus, schmittcreationfictional}, to information gathering~\cite{he2025deservecredit, long2023tweetorial}, drafting~\cite{park2021wrote, gero2022sparks}, and reviewing~\cite{mcfarland2024peerreview, lee2024design}. AI involvement may alter both the form and the content of the written text~\cite{sarkar2023exploring}, through editing writers' drafts to correct spelling and grammar errors~\cite{chang2015writeahead2, Fitria2021GrammarlyAA} and refining logic and flow to improve writing clarity~\cite{zhang2023visar, xia2022persua}. \CR{Through the lens of writing ecology~\cite{cooper1986ecology, prior2003chronotopic}, the continuous integration of AI in each granular step of the writing process may reconfigure the writer's metacognitive engagement with their own work.}


Writers' diverse practices in AI-assisted writing result in a wide range of interaction modes. Prior works introduced a number of frameworks and taxonomies to describe how writers used AI in their writing procedures (in other words, how AI was involved or contributed).~\citet{khosrowi2023diffusing} proposed the collective-centered creation (CCC) framework that described fine-grained creatorship in human-AI content co-creation according to five dimensions: relevance/(non-)redundancy and control, originality, time/effort, leadership and independence, and directness of each contributor's contribution in collaboration.~\citet{wan2024coco} proposed the CoCo matrix, which categorized writers' AI usage based on writers' and AI's cognitive contributions. They identified entropy and information gain as two dimensions and mapped writers' AI usage into four quadrants: high entropy and high information gain (e.g., writers plan, AI executes the plan), high entropy and low information gain (e.g., AI expands writers' drafts), low entropy and high information gain (e.g., AI plans, writers execute), low entropy and low information gain (e.g., AI provides feedback on writers' writings).~\citet{xupsychownership} categorized AI involvement in co-creation processes based on originality of contribution and level of contribution, where level of contribution can be further categorized based on sense of control and amount of effort in collaboration.~\citet{he2025deservecredit} described AI contribution based on the type of contribution, the amount of contribution, and the initiative. These frameworks and taxonomies share many commonalities (for example, many of them identified originality, effort, and initiative as dimensions to assess AI involvement), although they also diverge from one another. 
We synthesize these prior works and describe AI involvement as \emph{procedural factors} in Section~\ref{sec:factors_procedural}.

\subsection{Disclosure of contributors in collaborative content creation} 

\subsubsection{Disclosing and attributing to human collaborators} 

When people collaborate with other humans in content creation, it is common practice to disclose each contributor's work \CR{as an act of giving credit}~\cite{riley1996crafting}. In collaborative writing, this is commonly done by authorship attribution.\footnote{\CR{In the context of human-human co-creation, ``disclosure,'' ``attribution,'' and ``crediting'' are largely intertwined and often used interchangeably. In this paper, we use ``disclosure'' to describe the binary act of revealing other contributors to created content, whether humans or AI, whereas ``crediting'' is a motive for disclosure, and ``attribution'' is a common method of disclosure.}}
For example, in academic paper writing, researchers follow established rules to acknowledge each individual's effort and contribution~\cite{authorshipcontributiondisclosure}. The International Committee of Medical Journal Editors (ICMJE) sets a standard of authorship: making substantial contributions to the work, drafting or critically revising it, approving the final version, and being accountable for accuracy or integrity~\cite{icmjeauthorshipcontributors}. 
Similar standards extend to broader creative domains such as music and art.
For example, the Creative Commons (CC) formalizes collaborative authorship and reuse, setting rules for disclosure and attribution to the original creator(s) when the work is shared, reused, or modified~\cite{creativecommons}. 
In informal contexts, disclosing collaborators' contributions and authorship attribution are usually not required by explicit rules or policy; yet, disclosure is still considered a common practice due to social norms or moral considerations~\cite{youtubecollab, githubCodeOfConduct}. 

Humans give credit to collaborators to maintain fairness and adhere to moral norms, whereas misclaiming the full ownership of a collaborative work by not attributing credits to collaborators is disfavored~\cite{shaw2012children, pierce2003state}.
This is in line with the social exchange theory, where fair exchanges such as giving credit can enhance trust in social relationships and build social bonds~\cite{Cropanzano2017SocialET, Whitham2021GeneralizedGH}. Credit attribution is a core practice that fosters respect among creators and is essential in enhancing the authenticity and legitimacy of creative works~\cite{sinnreich2009ethics, scratchmanualcredit}. 

\subsubsection{Disclosing AI use and acknowledging AI's contributions: expectations and practices}

The rise of AI use in writing triggers discussions around whether or how AI's contributions to writing should be disclosed~\cite{burrus2024unmaskingAI}. Some platforms and communities have set up restrictive rules or policies about circumstances where AI contributions have to be disclosed~\cite{apnews2025amazonaidisclosure, lloyd2025redditAIrule, princeton2023generativeai}. 
In academic research, AI disclosure is usually mandatory when it is intentional and substantial ~\cite{resnik2025scientificresearchAIdisclosure}. In schools, some teachers clearly communicate AI use rules and give lower grades when undisclosed AI-generated content is detected~\cite{adnin2025eduperspective}. 

Other than policymakers of platforms or community regulators, readers, as content consumers, also show a trend of expecting transparency in AI use~\cite{cheong2025penalizingtransparencyaidisclosure, schilke2025transparencydilemma}. Trusting News surveyed over 6000 participants and found that 93.8\%  of the news readers hope journalists will disclose their AI use~\cite{newresearch}.
In a survey of AI use in brand advertising,  60\% of participants think disclosure is necessary in ``generating descriptions and taglines on what is being sold''~\cite{yougov2024aibrandads}.

However, we have not seen many writers proactively disclose their AI use.
~\citet{draxler2024ai} observed the AI ghostwriter effect, meaning that content creators are less likely to acknowledge contributions made by AI than by a human collaborator. 
~\citet{draxler2024ai} finds that writers are less likely to disclose AI contribution compared to a human collaborator's contribution in writing, which is conceptualized as the ``AI ghostwriting effect.'' Students strategically hide their AI use by modifying the tone and style of AI-generated content before incorporating it into their assignments~\cite{adnin2025eduperspective}. 
There are many reasons why AI use in writing is rarely disclosed. Many AI use policies do not provide enough guidance on how to disclose~\cite{weaver2024artificial, shuklahcistudent}. 
The fear of stigma and judgment around AI-assisted content creation also contributes to people's non-disclosure~\cite{giray2024AIshaming, he2025deservecredit, zhang2025secret, wang2024gpttabesl}. Writers fear that disclosing AI use will undermine the perceived originality, quality, and credibility of their work~\citet{he2025deservecredit}, or cause their capability being questioned and judged~\cite{zhang2025secret}. These concerns are reasonable---evidences show that when content is labeled as AI-generated or AI-edited, consumers tend to view it as less valuable, authentic, trustworthy, and of lower quality~\cite{li2024doesdisclosureaiassistance, schilke2025transparencydilemma, rubin2025perceivedempathy, kucinskas2025aiadcreation, horton2023biasaiart}, which is conceptualized as ``AI shaming''~\cite{giray2024AIshaming}.

%% file: factors_hypo.tex
\section{Factors and hypotheses} \label{sec:factors_and_hypo}

Previous studies identified many factors that potentially influence people's views on AI use in writing and their disclosure behaviors~\cite{he2025deservecredit, zhang2025secret, khosrowi2023diffusing, fang2025shapes}. Building on those studies, we synthesize a list of factors that may affect the perceived necessity of disclosing AI use, grouped into three dimensions: \emph{perspective} (Section~\ref{sec:factors_perspective}), \emph{purpose} (Section~\ref{sec:factors_purpose}), and \emph{procedural factors} (Section~\ref{sec:factors_procedural}). We describe these factors and pose hypotheses below.


\subsection{Perspective}
\label{sec:factors_perspective}

We investigate the perceived necessity of disclosure from the perspectives of two stakeholders who directly engage with AI-assisted writing and AI disclosure: readers and writers: 
\begin{itemize}
 \item \textbf{Readers} are individuals who consume or interpret written texts produced by writers.
 \item \textbf{Writers} are individuals who create or produce written texts intended to be shared with readers. 
\end{itemize}

Readers and writers may have different motivations for providing or seeking AI disclosure, and disclosure may have different influences on readers and writers. For example, writers may consider disclosing AI use in writing as a moral behavior, which aligns with their values and standards of honesty and transparency and helps build trust~\cite{liao2023ai, renieris2024artificial}. Meanwhile, writers and their works can potentially be negatively affected by disclosure, as people may perceive AI co-created writings to as less valuable or think writers who need AI assistance are less competent~\cite{li2024doesdisclosureaiassistance, cardon2025professionalism, jia2024news}. On the other hand, many readers expect AI involvement to be disclosed as they believe in the utility of disclosed information. For example, readers can assess how much they should trust the content or even decide whether the content is worth reading~\cite{fletcher2024does, gilardi2024willingness, toff2024or}. Thus, we hypothesize that the perspective has a main effect on the perceived necessity of AI disclosure (\textbf{H1}); compared to writers, readers deem disclosure to be more necessary (\textbf{H1a}).  



\subsection{Purpose}
\label{sec:factors_purpose}

The purpose of the written text reflects readers' and writers' goals of reading and writing. To select purposes to be studied in this work, we review HCI literature on AI-assisted writing and categorize the writing scenarios/tasks investigated in prior works based on the purpose of writing~\cite{he2025deservecredit, zhang2025secret, li2024doesdisclosureaiassistance, rae2024effects, hwang202580, hoque2024hallmark, adnin2025eduperspective,schilke2025transparencydilemma, resnik2025scientificresearchAIdisclosure, lee2024design, cardon2025professionalism, jia2024news}. We only consider the purposes of written texts intended for sharing with others (instead of diaries or notes kept for personal use), as AI disclosure is most relevant to such texts. We select three purposes representative of goals in common reading and writing scenarios, while recognizing that they are neither exhaustive nor mutually exclusive:


\begin{itemize}

 \item \textbf{Evaluation} refers to the cases where the purpose of the text is to enable the readers to evaluate the writer. Example written texts with this purpose include students' school applications and job applicants' cover letters.  
 \item \textbf{Learning} refers to the cases where the purpose of the text is to inform or teach, helping readers gain information, knowledge, or skill. Example written texts with this purpose include educational articles and technical tutorials.
 \item \textbf{Entertainment} refers to cases where the purpose of the text is to bring enjoyment to readers. Example written texts with this purpose include casual blog posts and fictional stories.
\end{itemize}

The different purposes of the written text may lead to variance in the importance of claiming or knowing the sources or contributors of the writing. For example, Trusting News reported that a majority of news readers want journalists to disclose their AI use, which is likely due to readers' need to calibrate trust and estimate the credibility of news content~\cite{newresearch, henestrosa2024journalmedia}. The perceived necessity of disclosure might be reduced in lower-stake reading or writing scenarios.
In addition, all three purposes may come with context-specific expectations and standards (e.g., whether AI use is acceptable in a specific context, and how likely the writer and the content will be judged and devalued)~\cite{zhang2025secret}. Such external judgments were found to significantly affect writers’ tendency to hide AI use~\cite{zhang2025secret}. Therefore, we hypothesize that the perceived necessity of disclosure differs based on the purpose of the written text (\textbf{H2}).

\subsection{Procedural factors} \label{sec:factors_procedural}

How writers use AI in the writing processes is a dimension to consider when it comes to AI disclosure~\cite{he2025deservecredit, khosrowi2023diffusing}, but how can we describe AI use in a finer-grained manner that goes beyond binary description (i.e., AI is or is not used)? As reviewed in Section~\ref{sec:background_cocreation}, prior works proposed multiple frameworks and taxonomies to describe AI use in writing (or more general content creation).
Although expressed using different terms, we find that they share significant overlap in meaning. Therefore, we synthesize constructs from prior works and distill them into a set of \emph{procedural factors}. 

Our procedural factors are drawn primarily from the collective-centered creation (CCC) theoretical framework---a comprehensive set of features of human-AI co-creation procedures---designed to inform how credit should be distributed~\cite{khosrowi2023diffusing}. 
Yet, the scope of the CCC framework goes beyond writing.
To better ground procedural factors in the context of writing, we borrowed ideas from recent studies about writers' AI usage, each touches on a subset of the CCC framework's features and offers empirical evidence on their effects (e.g.,~\cite{he2025deservecredit, xupsychownership}).
We synthesize these insights and define the four procedural factors (see the mapping from procedural factors to the sources in Appendix~\ref{appendix:procedural_factor_sources}). 
Note that although we use binary levels (high and low) to illustrate examples and simplify our vignette design (see Section~\ref{sec:method_vignette_design}), each procedural factor, by definition, represents a continuous spectrum: 

\begin{itemize}
    \item \textbf{Replaceability} denotes to what degree the writing could have been produced by the writer without AI (i.e., to what extent AI contributions in the writing were replaceable). High replaceability means that the writer could create a comparable work without AI, while low replaceability means that the writer is not capable of creating a similar writing if AI is taken away. 
    \item \textbf{Effortfulness} refers to how much the writer invested in writing. High effortfulness means that the writer put in much effort in writing with AI (e.g., spending considerable time on constructing and refining prompts), and low effortfulness means that the writer put in little effort when writing with AI (e.g., using AI to automate a large proportion of the work). 
    \item \textbf{Intentionality} describes to what degree the writer had a concrete blueprint of the writing in their mind and steered AI generation towards the goal intentionally and purposefully. High intentionality means that the writer used AI in a highly goal-oriented way (e.g., prompting AI to generate the specific content they want), while low intentionality means that the writer provided little guidance and let AI take the lead in content creation.  
    \item \textbf{Directness} denotes the extent to which AI contributions were precisely reflected in the writing. High directness means that the writer directly incorporated AI generation into their writing with minimal editing, while low directness means that the writer only used AI generation indirectly (e.g., using AI generation as a reference without copying it into their work). 
\end{itemize}

Different AI usage can lead to varied perceptions or psychological effects~\cite{he2025deservecredit, draxler2024ai, polimetla2025paradigm, khosrowi2023diffusing, giray2024AIshaming}. Readers may also judge whether AI involvement is worth disclosure based on the type and amount of AI contribution~\cite{he2025deservecredit}. Additionally, the extent of AI's influence on the writing outcome affects the writers' perceived ownership and agency~\cite{draxler2024ai, lim2025agency, xupsychownership, psychownershiplongprompt, biermanntoolcompanion}, which may shape their perceived necessity of disclosure. 
Informed by these works showing that AI involvement in writing may have impacts on readers' and writers' judgments and that high AI involvement may increase their tendency of disclosure, we hypothesize that the procedural factors have main effects on the perceived necessity of AI disclosure (\textbf{H3}). Furthermore, we hypothesize that low replaceability (\textbf{H3a}), low effortfulness (\textbf{H3b}), low intentionality (\textbf{H3c}), high directness (\textbf{H3d}) increases the perceived necessity of AI disclosure.


Some procedural factors may have different effects on readers and writers. For example, effortfulness could affect writers' perceived ownership of the writing~\cite{polimetla2025paradigm}, thus it might be more relevant to writers' perceived necessity of disclosure but not as important from the readers' perspective. Therefore, we hypothesize that the perspective moderates the effects of procedural factors (\textbf{H4}).

%% file: method.tex
\section{Method}





\subsection{Vignettes for research}

\subsubsection{Methodological considerations}

We treat all factors described in Section~\ref{sec:factors_and_hypo} as independent variables that may affect the perceived necessity of AI disclosure. This means that our study needs to include a large number of combinations of factors, and it is impractical to simulate empirical experiences that reflect all of them (e.g., by implementing many variations of prototypes that study participants can interact with). Furthermore, procedural factors are not manipulable in participants' lived experiences, as we cannot control how participants use AI to write.

To overcome these barriers, we choose to conduct a vignette experiment.
Vignette experiments use descriptions of hypothetical situations or contexts (or ``vignettes'') as probes to elicit respondents' perceptions, decisions, or reactions~\cite{payton2022vignette, atzmuller2010experimental}.
A vignette experiment allows us to (1) easily manipulate independent variables and systematically create combinations of variables~\cite{alexander1978use, atzmuller2010experimental}, (2) examine situations that are complicated or unethical to be replicated empirically~\cite{li2023assessing}, and (3) provide richer and more realistic contexts compared to traditional survey questions to obtain reliable attitudes and opinions from respondents and predict actual behavior~\cite{alexander1978use, wason2002designing, reno2016matters}. 
Thanks to these benefits, vignette experiments have been widely used in HCI (e.g.,~\cite{grgic2019human, 10.1145/3411763.3443433, 10.1145/3626705.3627766, 10.1145/3613904.3642042}).



\subsubsection{Vignette design} \label{sec:method_vignette_design}

We manipulate perspective, purpose, and procedural factors through our vignette design. Perspective includes two levels: the readers' perspective and the writers' perspective. Purpose includes three levels: evaluation, learning, and entertainment. To account for the natural variations within each purpose and enhance generalizability, we describe each purpose using two \emph{scenarios} that mirror real-world reading and writing settings. 
\CR{Each of the four procedural factors includes two levels: high and low. While these procedural factors are inherently continuous, we reduced them to a binary scale for simplicity. Together, the combinatorial variation across perspective, purpose, and the four procedural factors yields a diverse set of vignettes that may elicit divergent perceptions of the necessity of disclosure.
Fig.~\ref{fig:vignette_combined} (and textual version in Appendix~\ref{appendix:factor_operationalization} Table~\ref{table:factor_operationalization}) illustrates the vignette versions presented to the study participants, reflecting our operationalization of the factors introduced in Section~\ref{sec:factors_and_hypo}. }

\begin{figure}[htbp]
  \centering
  \includegraphics[width=\linewidth]{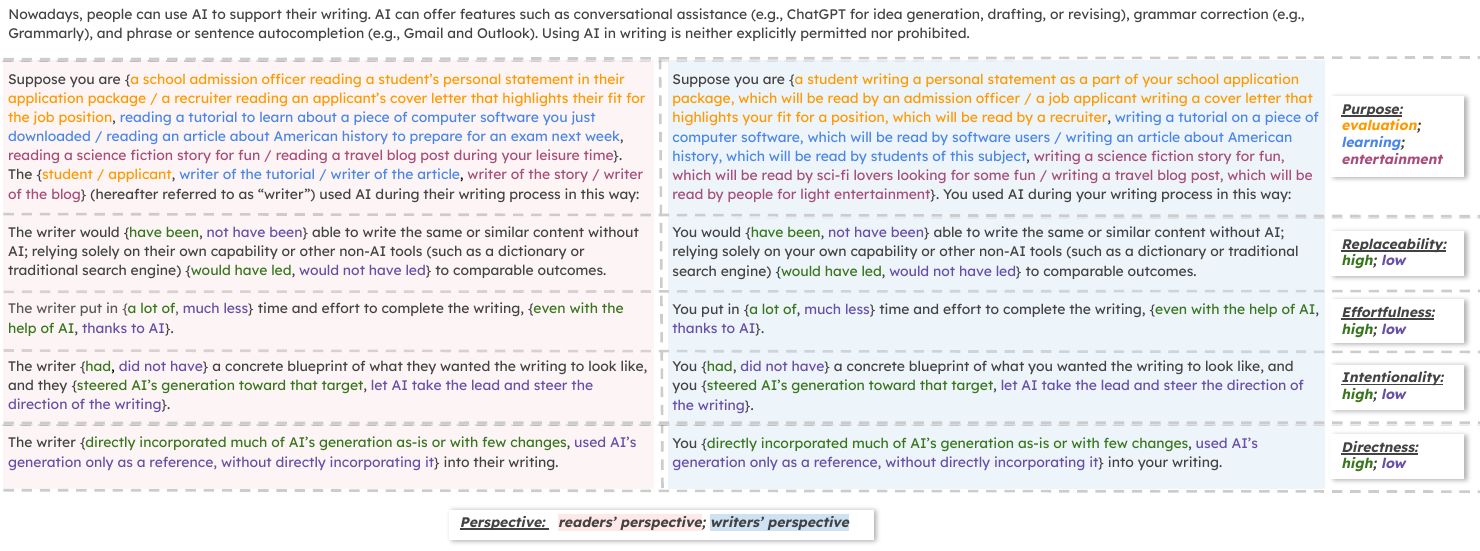}
  \caption{Vignettes from \colorbox{lightpink}{readers' perspective (left)} and \colorbox{lightblue}{writers' perspective (right)}. Both vignettes embed texts that manipulate purpose (\textcolor{orange}{evaluation}, \textcolor{darkblue}{learning}, or \textcolor{lightpurple}{entertainment}) and four procedural factors:  replaceability, effortfulness, intentionality, and directness (\textcolor{darkgreen}{high} or \textcolor{darkpurple}{low} for each procedural factor). }
  \Description{The figure shows the vignette from readers' and writers' perspectives. The vignette describes one reading or writing scenario according to the purpose assigned, and embeds one level for each procedural factor.}
  \label{fig:vignette_combined}
\end{figure}

Specifically, each vignette starts with describing the context of AI-supported writing and stating that AI use is not explicitly disallowed in writing, as we cannot put participants in a situation where they disobey the rule. We are studying how people perceive the necessity of disclosure, instead of judging whether AI can or should be used in writing. Each vignette embeds one perspective (out of the two perspectives), one scenario that reflects one purpose (out of the three purposes), and one level for each of the four procedural factors. Both perspective and scenario are randomly selected. We employ Resolution IV fractional factorial design to construct combinations of levels in procedural factors.\footnote{This approach reduces the total number of combinations in a full factorial design from $2^4$ to $2^{4-1}$. It allows us to reduce the number of participants and the cost required in the data collection without sacrificing the ability to identify significant factors. The disadvantage is that the main effects may be confounded with 2-factor interactions, which is not a problem because we do not consider interaction effects between procedural factors in our data analysis.} 

\subsection{Data collection}

We deployed the study in August 2025. The study is implemented in Qualtrics.\footnote{\url{https://www.qualtrics.com/} Accessed in 2025.} The data collection is approved by the authors' Institutional Review Board (IRB).

\subsubsection{Recruitment}

We recruited study participants from Prolific.\footnote{\url{https://www.prolific.com/} Accessed in 2025.} 
Participants who fulfill the following requirements are eligible to participate: (1) Above 18 years old; (2) Currently living in the United States; (3) Proficient in reading and writing in English; (4) Do not have reading disabilities or cognitive impairments that could affect decision-making ability; (5) Understand how AI can be used to support writing. These requirements are checked through participants' self-reported information in a screening questionnaire. We recruited 823 participants in total. Among them, 41 did not finish the study, were flagged as bot-like (reCAPTCHA score $<= 0.5$), or made duplicate submissions, and 55 did not pass the attention checks embedded in the study. We analyze data from 727 participants after filtering. The sample size is appropriate according to our simulation-based power analysis ($\alpha = 0.05$, power $> 0.80$, log odds $= -0.90$). 
We provide the participants' demographic information in Appendix~\ref{appendix:post-study-questionnaire}.

\subsubsection{Study procedure} \label{sec:method_study_procedure}

\begin{figure}[htbp]
  \centering \includegraphics[width=0.85\linewidth]{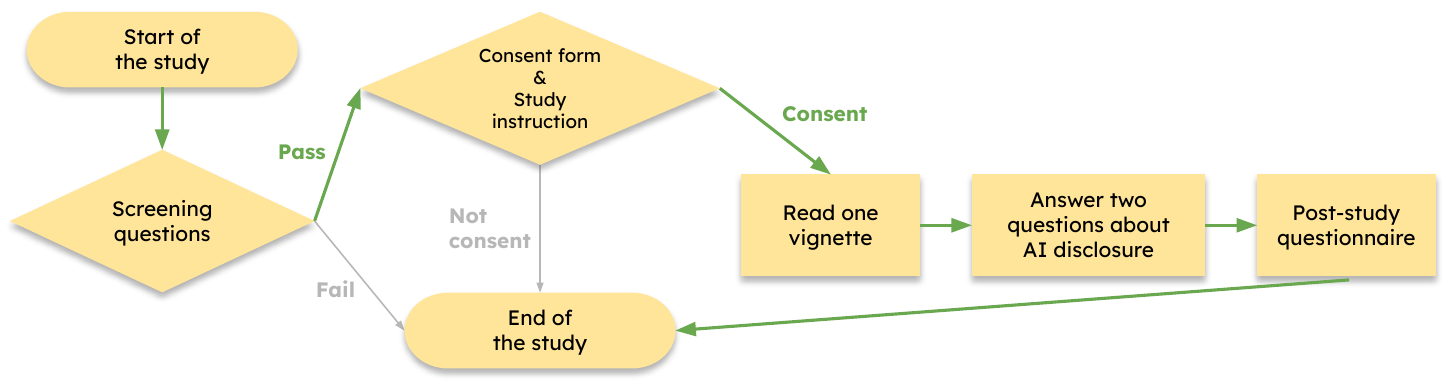}
  \caption{Study procedure. The green bolded arrows lead to study completion, while the gray arrows lead to quitting due to being ineligible or not consenting.}
  \Description{The figure shows a flowchart indicating the procedure of the study. Each participant completes the following steps: answering screening questions, reading the consent form and instruction, reading one vignette, answering questions about the perceived necessity of disclosure, and filling out the post-study questionnaire. }
  \label{fig:study_procedure}
\end{figure}

Fig.~\ref{fig:study_procedure} visualizes the study procedure. Each participant begins by answering screening questions that verify their eligibility for participation. They are shown the consent form and study instruction if they pass the screening. If the participant consents, they are presented with a vignette that outlines a reading or writing situation as described in Section~\ref{sec:method_vignette_design}. Based on the vignette, each participant indicates whether they perceive AI disclosure as necessary (choose from binary Yes/No). We opt to ask a binary choice question, rather than a Likert-scale question that allows more subtle opinions (e.g., ``disclosure is somewhat necessary'') because  (1) the behaviors that their perceptions ultimately lead to (requesting or providing AI disclosure or not) are inherently binary and (2) we seek to avoid the central tendency bias~\cite{stevens1971issues, douven2018bayesian}. Each participant then explains their choice in an open-ended response.

As the final step, each participant fills out a post-study questionnaire. In addition to standard demographic questions, the questionnaire also includes 13 items to capture four personal factors: (1) \emph{Self-efficacy of writing} refers to a person's belief in their own capability of writing~\cite{bandura1977self}. Some prior work showed intricate relationships between one's AI usage and self-efficacy~\cite{Lai2025aiblendedlearningwriting, kim2024jobinsecurity}. 
(2) \emph{AI literacy} refers to a person's competencies in understanding, evaluating, and using AI~\cite{long2020ai}, which could influence one's behavior and perception when using AI for writing. (3) \emph{Internal judgment about AI use} refers to a person's self-judgment when using AI, such as whether using AI in a certain context is appropriate or moral, which can also affect their behaviors during AI use~\cite{liang2025widespreadadoptionlargelanguage}. (4) \emph{Internal judgment about disclosure} refers to a person's self-judgment of (non-)disclosure of information. For example, AI disclosure is usually considered a practice reflecting one's value of transparency and honesty~\cite{zhang2025secret}.
We present all questions in the post-study questionnaire in Appendix~\ref{appendix:post-study-questionnaire}. The study takes about 10 minutes, and participants are paid \$2.00 for their participation.


\subsection{Data analysis}


We build a logistic regression model with participants' binary choices of perceived necessity of AI disclosure as the dependent variable. To test our hypotheses about the main effects (H1-H3), we include perspective, purpose, replaceability, effortfulness, intentionality, and directness as independent variables. To study the moderation effect in H4, we add the interaction terms between perspective and each procedural factor.\footnote{For sanity check, we built another logistic regression model with all two-factor interaction terms between purpose and procedural factors, in addition to the independent variables in the reported model. We picked the reported model because it is more interpretable and yields higher statistical power. We confirm that the model we pick balances goodness of fit and simplicity (evidenced by its lower AIC).} 
We include the four personal factors from Section~\ref{sec:method_study_procedure} as covariates to account for individual differences that may influence the perceived necessity of AI disclosure. 

%

We also conduct a supplementary qualitative analysis of participants' open-ended responses, which is not a part of the main study. We report the analysis procedure in Appendix~\ref{appendix:method_qual} and the findings in Appendix~\ref{appendix:results_qual}. 




%% file: results.tex
\section{Results} \label{sec:results}

\begin{figure}[htbp]
  \centering
  \begin{minipage}[t]{0.63\textwidth}
    \centering
    \small
    \begin{tabular}{p{0.65\textwidth} r l}
      \toprule
      \textbf{} & \shortstack{\textbf{Model} \\ Coef. (S.E.)}  & \shortstack{\textbf{Sig.} \\ {} } \\
      \hline
      (Intercept) & -4.838 (1.118) & *** \\
      \multicolumn{2}{l}{\textit{\textbf{Main effects}}} \\
      \rowcolor{blue!15}
      \quad Perspective: writers' vs. readers' & -1.593 (0.455) & *** \\
      \quad Purpose: learning vs. evaluation & 0.231 (0.241) & n.s.\\
      \quad Purpose: entertainment vs. evaluation & 0.219 (0.237) & n.s.\\
      \rowcolor{blue!15}
      \quad Replaceability: high vs. low & -0.756 (0.279) &  **\\
      \quad Effortfulness: high vs. low  & -0.055 (0.275) & n.s.\\
      \rowcolor{blue!15}
      \quad Intentionality: high vs. low & -1.239 (0.282) & *** \\
      \rowcolor{blue!15}
      \quad Directness: high vs. low & 0.939 (0.276) & ** \\
      \multicolumn{2}{l}{\textit{\textbf{Interaction terms}}} \\
      \quad Replaceability $\times$ Writers' perspective & 0.447 (0.399) & n.s.\\
      \quad Effortfulness $\times$ Writers' perspective & -0.177 (0.392) & n.s.\\
      \rowcolor{blue!15}
      \quad Intentionality $\times$ Writers' perspective & 1.547 (0.397) & *** \\
      \quad Directness $\times$ Writers' perspective & -0.119 (0.392) & n.s.\\
      \multicolumn{2}{l}{\textit{\textbf{Covariates}}} \\
      \quad Writing self-efficacy & -0.367 (0.161) & * \\
      \quad AI Literacy & -0.518 (0.190) & ** \\
      \quad Internal judgment of AI use & 0.209 (0.119) & $^{+}$ \\
      \quad Internal judgment of disclosure & 2.139 (0.185) & *** \\
      \hline
      Num. Obs. & 727 \\
      Pseudo $R^2$ & 0.352 \\
      \bottomrule
      \multicolumn{2}{l}{\footnotesize{Note: + $p < 0.1$, * $p < 0.05$, ** $p < 0.01$, *** $p < 0.001$}}
    \end{tabular}
    \captionof{table}{Logistic regression model predicting the perceived necessity of AI disclosure in writing. This table presents coefficients as log odds, standard errors (in parentheses next to the coefficients), and statistical significance (annotated with stars). In the main effects section, the second level in each comparison pair is the reference level (i.e., reference level = reader' perspective, evaluation purpose, low replaceability, low effortfulness, low intentionality, low directness). A positive coefficient means \emph{increased} perceived necessity compared to the reference level. The table shows that perspective (readers' vs. writers'), replaceability (high vs. low), intentionality (high vs. low), and directness (high vs. low) have main effects on the perceived necessity of disclosure. 
    The interaction between intentionality and perspective is also significant. 
    }
    \Description{This table shows the logistic regression model presenting the effects of perspective, purpose, procedural factors, the interaction terms between perspective and procedural factors, and the covariates. For each term, we present log-odds, standard errors, and p-values.}
    \label{table:logit}
  \end{minipage}%
  \hfill
  \begin{minipage}[t]{0.35\textwidth}
    \centering
    \vspace{-4.5cm}
    \includegraphics[width=\linewidth]{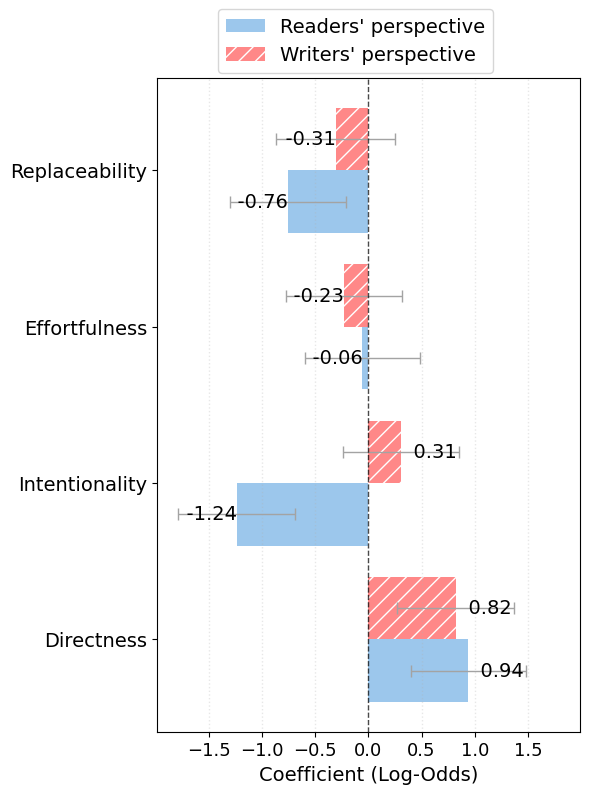}
    \caption{A bar graph showing the procedural factors' differing effects on perceived necessity of AI disclosure based on perspective, according to the logistic regression model shown in Table~\ref{table:logit}. 
    A positive coefficient means that a high-level of the procedural factor (e.g., high directness) \emph{increases} the perceived necessity of disclosure compared to the procedural factor's reference level (low-level); a negative coefficient means that a high-level of the procedural factor (e.g., high replaceability) \emph{decreases} the perceived necessity of disclosure compared to the procedural factor's reference level (low-level).
    }
    \Description{The figure shows a bar graph comparing the effect of each procedural factor based on perspective. }
    \label{fig:result_interaction}
  \end{minipage}
\end{figure}

We report the findings from the logistic regression. Table~\ref{table:logit} presents the effects of perspective, purpose, and procedural factors on the perceived necessity of AI disclosure in writing, with the coefficients (log odds), p-values, and standard errors in the logistic regression model. 
We reflect on these findings and contextualize them with respect to prior works in Section~\ref{sec:discussion_findings}.

\paragraph{\underline{Perspective (RQ1)}}

There is a statistically significant difference in the perceived necessity of disclosure from readers' and writers' perspectives. \textbf{Compared to readers, writers are significantly less likely to perceive AI disclosure as necessary} (Coef.=-1.593, p<0.001, 95\% CI: [-2.485, -0.700]). In other words, writers have 79.7\% lower odds of perceiving disclosure as necessary compared to readers. H1 and H1a are supported. 

\paragraph{\underline{Purpose of the written text (RQ2)}}


We see a trend that readers and writers are more likely to perceive disclosure as necessary when the purpose of the written text is learning rather than evaluation (Coef.=0.231, p=0.337, 95\% CI: [-0.240, 0.703]), but the difference is not statistically significant.
Similarly, when the purpose of the written text is entertainment, readers and writers show a non-significant tendency to perceive disclosure as more necessary than when the purpose is evaluation (Coef.=0.219, p=0.355, 95\% CI: [-0.245, 0.683]). 
In summary, we find no evidence that the purpose of the written text affects the perceived necessity of AI disclosure. H2 is not supported.

\paragraph{\underline{Procedural factors (RQ3)}}

We found that replaceability has a main effect on the perceived necessity of disclosure. 
\textbf{Disclosure is significantly more likely to be perceived as necessary when writers cannot easily replace AI contributions in the writing} (i.e., low replaceability) (Coef.=-0.755, p<0.01, 95\%CI:[-1.302, -0.209]).
H3a is supported. Surprisingly, the effect of effortfulness on perceived necessity of disclosure is not significant (Coef.=-0.055, p=0.841, 95\%CI: [-0.595, 0.484]). H3b is not supported. Intentionality has a main effect on the perceived necessity of disclosure. \textbf{Disclosure is significantly more likely to be perceived as necessary when writers do not take the lead and intentionally guide AI generation} (i.e., low intentionality) (Coef.=-1.239, p<0.001, 95\%CI:[-1.792, -0.686]). H3c is supported. We also found that directness has a main effect on the perceived necessity of disclosure. \textbf{Disclosure is significantly more likely to be perceived as necessary when writers directly incorporated AI contributions into the finalized written text as-is or with few changes} (i.e., high directness) (Coef.=0.9391, p=0.001, 95\%CI: [0.398, 1.480]). H3d is supported. 


The interaction between perspective and intentionality is found to be significant, indicating that \textbf{readers and writers diverge in how intentionality shapes their perceived necessity of disclosure}. 
Counterintuitively, when AI generation was intentionally steered towards a specific goal (i.e., high intentionality), readers perceive disclosure as less necessary; in contrast, writers perceive disclosure as more necessary (Coef.=1.547, p<0.001, 95\%CI: [0.769, 2.326]). 
The interaction effects between perspective and the other three procedural factors (i.e., replaceability, effortfulness, and directness) are not statistically significant. H4 is partially supported. In Fig.~\ref{fig:result_interaction}, we visualize the coefficients of procedural factors' main effects and the interaction terms in the logistic regression model (Table~\ref{table:logit}), showing how the effects of the procedural factors on the perceived necessity of AI disclosure vary based on the perspective taken.

\paragraph{\underline{Covariates}}
As expected, several covariates regarding personal background and values also show significant associations with their perceived necessity of disclosure. We found that participants with high writing self-efficacy and AI literacy are less likely to perceive disclosure as necessary. In contrast, participants who have a strong internal judgment of disclosure (capturing how much they value transparency and honesty) are more likely to consider disclosure necessary. 
We note that these covariates are included primarily as controls and are not robustly manipulated as independent variables. While the significance of covariates suggests that individual differences may play a role in the perception of disclosure necessity, we lack sufficient power to draw a conclusion about the covariates' effects on the perceived necessity of disclosure, and the related findings should be interpreted as exploratory.

%% file: discussion.tex
\section{Discussion}


\subsection{Reflection on study findings} \label{sec:discussion_findings}



Our study result shows a substantial gap in the perceived necessity of AI disclosure between perspectives, indicating that writers are much less likely to perceive disclosure as necessary compared to readers. 
This gap may be explained by the \emph{self-serving bias}, which suggests that people tend to attribute desirable outcomes to their own capabilities or behaviors~\cite{shepperd2008exploring, miller1975self}. In the context of this study, self-serving bias may cause writers to overly ascribe the outcome of AI-assisted writing to themselves, thereby underestimating AI's contribution and consequently perceiving a reduced need for disclosure. \CR{Furthermore, this reader-writer gap may be tied to the pervasiveness of AI in the writing ecology~\cite{cooper1986ecology}. As AI becomes seamlessly integrated into writing interfaces and platforms today, writers may perceive it as a form of infrastructure that naturally exists within the social environment (or in other words, an ``ambient environmental factor'')~\cite{prior2003chronotopic, star1994steps}. This diminished visibility of AI from the writers' view may contribute to their lower perceived necessity of disclosure.}
As expected, high AI involvement in writing generally increases the perceived necessity of AI disclosure. However, effortfulness (i.e., how much effort a writer puts into writing) does not have a notable impact on the perceived necessity of disclosing AI use. This appears to contradict the \emph{effort heuristic}, a well-known effect in cognitive psychology, which demonstrates that people assign value to a work based on the effort it took to produce it~\cite{kruger2004effort, ziano2023effort}. Following the effort heuristic, one would expect that the greater effort invested by a writer should increase the perceived value of the writer's contribution, which in turn reduces the value assigned to AI contribution and lessens the perceived necessity of disclosure. 
We also initially hypothesized that the purpose of the written text may affect readers' and writers' judgment about whether disclosure is necessary, because readers and writers appear to care more about AI use in some writing scenarios than others (e.g., AI use seems to be more acceptable in creative writing~\cite{hwang202580} than in writing journalistic content~\cite{toff2024or}).
While our finding of the insignificance of purpose was initially surprising, we later found that it loosely aligns with the findings from prior work. \citet{he2025deservecredit} found that writing context (writing in academic, professional, or technical context) is not a significant predictor of how writers credit authorship to AI, which may correlate with their perceived necessity of disclosure. On the other hand, our supplementary qualitative analysis shows that for some participants, purposes of reading and writing do have an impact on their perceptions of AI disclosure necessity (see Appendix~\ref{sec: qual_purpose}). Future work can further investigate this counterintuitive result on the effect of purpose.

\subsection{Considerations on the future design of AI disclosure guidance and tools}


The readers' and writers' differing perceptions of AI disclosure necessity underscore the importance of guidance and interventions that raise writers’ recognition of the need to disclose.
We note that many writers are aware of the benefits of disclosure (e.g., for meeting ethical standards and building trust, as evidenced by writers' open-ended responses reported in Appendix~\ref{appendix:results_qual}), suggesting that they do not hold an entrenched negative attitude towards AI disclosure and that motivating them to disclose AI use is feasible. 

\begin{figure}[htbp]
  \centering
  \includegraphics[width=\linewidth]{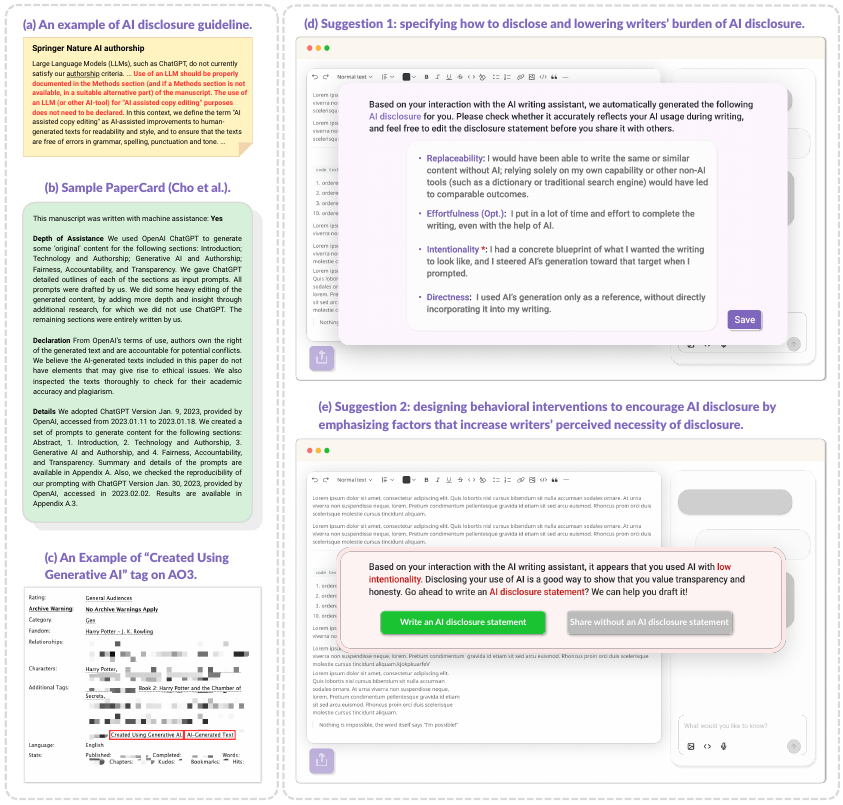}
  \caption{On the \emph{left}, we show three existing approaches that guide writers' AI disclosure. Part \textcolor{amethyst}{\textbf{(a)}} shows Springer Nature AI authorship policy, with the text directly coming from~\citet{springernaturepolicy}. Part \textcolor{amethyst}{\textbf{(b)}} shows a sample AI disclosure statement structured based on the PaperCard framework, with the text directly coming from~\citet{cho2023papercard}. Part \textcolor{amethyst}{\textbf{(c)}} shows an example of ``Created Using Generative AI'' tag (and its variant, ``AI-Generated Text'') on a content sharing platform, AO3, with a screenshot partially blurred for anonymization~\cite{ao3policy}. On the \emph{right}, we show two future directions for designing AI disclosure guidance and tools. Part \textcolor{amethyst}{\textbf{(d)}} demonstrates an envisioned tool that automatically constructs an editable granular AI disclosure statement (deemphasizes effortfulness by making it optional and emphasizes intentionality with a red asterisk), following a design suggestion in Section~\ref{sec: discussion_suggestion1}. Part \textcolor{amethyst}{\textbf{(e)}} demonstrates an envisioned nudging message as a behavioral intervention, which pops up when a writer attempts to share their writing without an attached AI disclosure statement, following a design suggestion in Section~\ref{sec: discussion_suggestion2}. }
  \Description{The figure shows three examples of existing approaches to enforce or encourage AI disclosure (through implementing a publication policy, through providing an AI disclosure framework, and through an optional ``Created Using Generative AI'' tag). It also shows two example prototypes that demonstrate our design suggestions provided in Section~\ref{sec: discussion_suggestion1} and Section~\ref{sec: discussion_suggestion2}, namely, specifying how to disclose and lowering writers' burden of AI disclosure, and designing behavioral interventions to encourage AI disclosure.}
  \label{fig:vignette_combined}
\end{figure}

Currently, most content-sharing platforms, publishers, and communities follow one of the two approaches to raise writers' perceived necessity of AI disclosure:
(1) formulating regulations that \textit{enforce} writers to disclose their AI use; for example, policies of ACM Publications and Springer Nature (Fig.~\ref{fig:vignette_combined}(a)) require full AI disclosure and ask writers to document their AI use in manuscripts~\cite{acmpolicy, springernaturepolicy}; (2) designing behavioral interventions that \textit{encourage} writers to disclose; for instance, 
Medium incentivizes AI disclosure by limiting the distribution and monetization of undisclosed AI-generated writing,
and AO3 (Fig.~\ref{fig:vignette_combined}(c)) promotes disclosure by prompting authors to label their AI-assisted work with a ``Created Using Generative AI'' tag~\cite{ao3policy}. 
Extending from these two approaches, we discuss insights into designing future AI disclosure guidance and tools grounded in our study findings.

\subsubsection{Scaffolding AI disclosure} \label{sec: discussion_suggestion1}


Sometimes writers do not disclose their AI use due to practical difficulties; for example, many writers do not know when AI disclosure is expected, what it should look like, and how to disclose~\cite{ganjavi2023bibliometric, weaver2024artificial}. Designers and regulators of content-sharing platforms could mitigate these obstacles and scaffold writers' AI disclosure.

Going beyond simply mentioning the disclosure requirement to offering more specific guidance (e.g., concrete examples or templates for disclosure) is one way to scaffold AI disclosure~\cite{ganjavi2023bibliometric, weaver2024artificial, cho2023papercard, hoque2024hallmark}.
Our study shows that readers and writers judge the necessity of disclosure based on \emph{how} AI contributes to the writing process, not merely \emph{whether} AI is involved, indicating that a binary disclosure statement or label is not sufficient (consistent with what~\citet{he2025deservecredit} suggested). 
Providing templates or exemplars that showcase a more granular description of AI use may guide writers to disclose more effectively by highlighting the specific details readers care about.
\CR{Writers could also construct their disclosure statements by describing their AI use along the four procedural factors we synthesized: the replaceability of AI,  the effort they invested, the intentionality of their AI use, and the extent to which AI-generated output was directly incorporated into writing. This paper's Generative AI usage statement shows an example.}

A shortcoming of scaffolding AI disclosure by providing disclosure templates is that writers still need to make a considerable effort to fill them out. 
To further lower the writers' burden, platforms can provide enhanced AI disclosure scaffolding by implementing features that partially automate the process of creating disclosure statements.
For example, by tracking the interactions between a writer and AI~\cite{lee2022coauthor}, platforms can estimate the writer's AI usage (e.g., whether the writer directly pasted AI-generated text with minimal edits), automatically construct a draft of a disclosure statement, and allow the writer to modify the draft before sharing it with readers. 
In Fig.~\ref{fig:vignette_combined}(d), we showcase how such an AI disclosure tool may look. Notably, it presents an automatically generated AI disclosure draft structured based on the four procedural factors. Effortfulness is labeled as optional, as our study finding shows that effortfulness does not affect readers' and writers' perceptions much; intentionality is emphasized with an asterisk, as the significant interaction effect between perspective and intentionality suggests that writers themselves may overlook the importance of disclosing their intentionality during AI-assisted writing. 

\subsubsection{Designing behavioral interventions to encourage AI disclosure by emphasizing factors that increase writers' perceived necessity of disclosure.} 
\label{sec: discussion_suggestion2}

Since some factors, such as individuals' value in transparency and honesty, increase writers' judgments on the necessity of disclosure, we envision that platforms could design nudging messages based on such factors as a behavioral intervention that encourages writers' proactive AI disclosure. Nudging messages gently guide people and alter their behaviors without restricting individual actions~\cite{caraban201923, thaler2021nudge}. For example, presenting a simple message such as ``Disclosing your use of AI is a good way to show that you value transparency and honesty.'' before a writer shares their work can remind them of the importance of disclosure while not undermining their agency and control. 

The nudging messages can be dynamically presented based on the writer's AI usage. \CR{By looking at the four procedural factors' different effects on readers' and writers' perceived necessity of AI disclosure, we can infer when such nudging is especially needed to get writers to disclose AI use and align with readers' expectations.} 
Our finding indicates that, from the readers' point of view, low intentionality (i.e., a writer lets AI take the lead in writing without intentionally steering AI generation) is the most important factor that necessitates AI disclosure; meanwhile, writers are more likely to judge disclosure as unnecessary when AI is used in a less intentional manner. Therefore, platforms may strongly nudge writers to disclose when low intentionality of AI use is detected, while applying weaker nudges when intentionality is high. We demonstrate this idea in Fig.~\ref{fig:vignette_combined}(e), which presents a nudging message that pops up when a writer attempts to share their writing without AI disclosure.

\subsection{Limitations}

There are several limitations of this work. Some of them come from the inherent methodological limitation of the vignette experiment. The vignettes require participants to imagine that they are readers and writers; future work can study the perceived necessity of disclosure in a more realistic setting where participants are actual readers and writers.
Relatedly, some participants may not be able to relate to the vignettes that they are assigned: people who seldom write outside of school or work contexts might find ``writing for entertainment'' hardly relatable, distorting their perceptions of the disclosure necessity. We try to minimize this problem by choosing reading and writing purposes that most people are familiar with. For participants who are assigned writers' perspective, their perceptions may be affected by social desirability bias~\cite{grimm2010social}, especially because disclosure is associated with moral considerations. Our result shows that writers are less likely to deem disclosure as necessary, even though writers might lean toward selecting ``Yes'' to appear to be ``in good faith'' in the vignette study. Thus, the observed difference between perspectives should be seen as a conservative estimate, and we expect the distortion caused by the social desirability bias would strengthen, rather than weaken, the reported reader-writer gap. Additionally, our study investigates the perception gap of AI disclosure necessity between readers and writers; our study findings cannot, and are not intended to, reflect the perception-action gap between readers' perceived necessity and writers' actual disclosure behaviors.

\CR{We aimed to improve generalizability by selecting two representative scenarios per purpose, but we acknowledge that these scenarios do not capture all possible variations within each purpose.}
\CR{While our vignettes explicitly operationalized procedural factors as high or low, their actual degrees are inherently subjective when readers attempt to assess based on the writer-AI interaction process during writing. In practice, people may apply different standards when judging the replaceability, effortfulness, intentionality, and directness of a writer's AI use. Among these factors, writers are better positioned to gauge replaceability (since writers have firsthand knowledge of whether they could have produced similar work without AI), whereas readers have no reliable way to assess it unless explicitly informed by the writers.}


To run the logistic regression, we have to pre-define and select a list of independent variables that may affect the perceived necessity of disclosure. Our supplementary qualitative analysis surfaced more factors that could change readers' and writers' perceived necessity of disclosure (see Appendix~\ref{appendix:results_qual}). \CR{Future work can treat these additional factors (such as social bias and stigma toward writings co-created with AI) as independent variables to investigate their influence on readers’ and writers’ perceived necessity of AI disclosure.}

%% file: conclusion.tex
\section{Conclusion}

AI-assisted writing has become a common part of contemporary writing practices. While AI disclosure is regulated in some cases, it is unclear what leads readers and writers to perceive disclosure as necessary. In this work, we present a vignette study that investigates how the perceived necessity of AI disclosure is influenced by perspectives, purposes, and procedural factors. 
We found a significant gap between perspectives: readers are more likely to consider disclosure to be necessary than writers. High AI involvement in writing (specifically, low replaceability of AI contribution, low intentionality of writers' AI use, and highly direct adoption of AI contribution in writing) increases the perceived necessity of AI disclosure. 
Perceived necessity of AI disclosure is additionally influenced by the interaction between perspective and intentionality of AI use: low intentionality increases readers’
perceived necessity but decreases writers’ perceived necessity. 
Our study takes a step towards understanding readers' and writers' perceptions around AI disclosure and suggests promising approaches for scaffolding AI disclosure and behavioral interventions. 

%% file: endmatter.tex
\begin{acks}
    \CR{We thank Ryan Carlson, Youngjae Cha, Inyoung Cheong, James Evans, Dongyoung Go, and Alex Shaw for their insightful discussions regarding the study design and data analysis. We thank the anonymous reviewers for their thoughtful feedback and our study participants for their valuable input.}
\end{acks}

\section*{Generative AI usage statement}

Using the concept of procedural factors described in this paper, AI was used with high replaceability, high effortfulness, high intentionality, and low directness.
Specifically, Grammarly's browser extension was used for grammar checking throughout the paper.\footnote{\url{https://www.grammarly.com/browser}. Accessed December 2025.} ChatGPT was occasionally used to improve sentence fluency (with the authors' original sentences provided via chat prompts), followed by the authors' careful review and extensive editing.\footnote{\url{https://chatgpt.com/}. Accessed December 2025.} ChatGPT was also used to improve the formatting of Table~\ref{table:procedural_factor_sources}.

%% file: appendix.tex
\clearpage

\appendix

\section{Sources of procedural factors} \label{appendix:procedural_factor_sources}

Table~\ref{table:procedural_factor_sources} shows and compares the sources of the procedural factors. \CR{The mapping from prior works to our procedural factors is based on high conceptual similarity and relevance rather than exact identicalness. For example, our definition of ``intentionality'' largely overlaps with ``leadership'' as described by Rezwana and Maher (i.e., whether AI leads or follows the creative process)~\cite{rezwana2023user}, while also mirroring the concept of ``entropy'' described by Wan et al. (e.g., high entropy describes writing processes that are started with a plan or base provided by the writers)~\cite{wan2024coco}. }

\begin{table}[htbp]
\centering

\begin{tabular}{@{}p{2.3cm}
                p{2.1cm}
                p{2.1cm}
                p{2.1cm}
                @{\hspace{0.5cm}}p{2.1cm}
                p{2.1cm}@{}}

\toprule

\textbf{} & \multicolumn{5}{c}{\textbf{SOURCES}} \\
\cmidrule(lr){2-6}
{\textbf{PROC. \newline FACTORS}} &
\textit{\citet{khosrowi2023diffusing}} &
\textit{\citet{he2025deservecredit}} &
\textit{\citet{rezwana2023user}} &
\textit{\citet{xupsychownership}} &
\textit{\citet{wan2024coco}} \\
\midrule

\textbf{Replaceability} &
Relevance/(non-)redundancy and control &
Contribution type &
User expertise &
Originality of co-creation &
\\
\hline

\textbf{Effortfulness} &
Time/effort &
Contribution amount &
 &
Amount of effort &
\\
\hline

\textbf{Intentionality} &
Leadership &
 &
Leadership &
Control in process &
Entropy \\
\hline

\textbf{Directness} &
Directness &
Initiative &
 &
 &
Information gain \\
\bottomrule

\end{tabular}

\caption{Mapping procedural factors to sources. Each row represents one procedural factor as defined in this paper, and how it was named/categorized in prior works. We note that prior works use different terms to describe the same or highly similar concepts.}
\label{table:procedural_factor_sources}
\end{table}

\section{Factor operationalization} ~\label{appendix:factor_operationalization}

Table~\ref{table:factor_operationalization} shows how we operationalize the factors that we investigate in this study. We include these operationalizations in our vignettes to describe perspective, purpose, and procedural factors.

\setlength{\tabcolsep}{7pt}   
\small                    

\begin{longtable}{>{\centering\arraybackslash}m{3em} >{\centering\arraybackslash}m{6em} p{39em}}

\caption{Factor operationalization showing how different levels are represented in the vignettes.} \label{table:factor_operationalization} 
\Description{In this table, we list the factors investigated in our study (Left), the levels included for each factor (Middle), and the operationalization reflecting how each level is represented in the vignettes presented to the participants (Right).}
\\

\toprule
\textbf{FACTORS} & \textbf{LEVELS} & \textbf{OPERATIONALIZATION} \\
\midrule
\endfirsthead

\multicolumn{3}{c}%
{{\bfseries Table \thetable\ (continued)}} \\
\toprule
\textbf{FACTORS} & \textbf{LEVELS} & \textbf{OPERATIONALIZATION} \\
\midrule
\endhead

\bottomrule
\endfoot

\bottomrule
\endlastfoot

\multirow{4}{*}{\rotatebox{90}{\makecell{\colorbox{Goldenrod}{\footnotesize{PERSPECTIVE}}}}}
& \multirow{2}{*} {\makecell{Reader}} & \rule{0pt}{3ex}Assign perspective \textcolor{blue}{[R]}, which is integrated in the operationalization of purpose and procedural factors.  \vspace{0.5em} \\
\cmidrule(lr){2-3}
& \multirow{2}{*} {\makecell{Writer}} & \rule{0pt}{3ex}Assign perspective \textcolor{orange}{[W]}, which is integrated in the operationalization of purpose and procedural factors.  \vspace{0.5em} \\
\midrule
\pagebreak  
\multirow{20}{*}{\rotatebox{90}{\makecell{\colorbox{Goldenrod}{\footnotesize{PURPOSE}}}}}
& \multirow{8}{*}{\makecell{Evaluation}} & \textcolor{blue}{[R]}  \textit{``Suppose you are a school admission officer reading a student's personal statement in their application package.''} \\ 
& & \textcolor{orange}{[W]} \textit{``Suppose you are a student writing a personal statement as a part of your school application package, which will be read by an admission officer.''} \\
\cmidrule(lr){3-3}
& & \textcolor{blue}{[R]} \textit{``Suppose you are a recruiter reading an applicant's cover letter that highlights their fit for the job position.''} \\
& & \textcolor{orange}{[W]} \textit{``Suppose you are a job applicant writing a cover letter that highlights your fit for a position, which will be read by a recruiter.''} \\
\cmidrule(lr){2-3}
& \multirow{8}{*}{\makecell{Learning}} & \textcolor{blue}{[R]} \textit{``Suppose you are reading a tutorial to learn about a piece of computer software you just downloaded.''} \\
& & \textcolor{orange}{[W]} \textit{``Suppose you are writing a tutorial on a piece of computer software, which will be read by software users.''} \\
\cmidrule(lr){3-3}
& & \textcolor{blue}{[R]} \textit{``Suppose you are reading an article about American history to prepare for an exam next week.''} \\
& & \textcolor{orange}{[W]} \textit{``Suppose you are writing an article about American history, which will be read by students of this subject.''} \\
\cmidrule(lr){2-3}
& \multirow{5}{*}{\makecell{Entertainment}} & \textcolor{blue}{[R]} \textit{``Suppose you are reading a science fiction story for fun.''} \\
& & \textcolor{orange}{[W]} \textit{``Suppose you are writing a science fiction story for fun, which will be read by sci-fi lovers looking for some fun.''} \\
\cmidrule(lr){3-3}
& & \textcolor{blue}{[R]} \textit{``Suppose you are reading a travel blog post during your leisure time.''} \\
& & \textcolor{orange}{[W]} \textit{``Suppose you are writing a travel blog post, which will be read by people for light entertainment.''} \\
\midrule

\multirow{12}{*}{\rotatebox{90}{\makecell{\colorbox{Goldenrod}{\footnotesize{REPLACEABILITY}}}}}
& \multirow{6}{*}{\makecell{Low}} & \textcolor{blue}{[R]} \textit{``The writer would have been able to write the same or similar content without AI; relying solely on their own capability or other non-AI tools (such as a dictionary or traditional search engine) would have led to comparable outcomes.''} \\
& & \textcolor{orange}{[W]} \textit{``You would have been able to write the same or similar content without AI; relying solely on your own capability or other non-AI tools (such as a dictionary or traditional search engine) would have led to comparable outcomes.''} \\
\cmidrule(lr){2-3}
& \multirow{6}{*}{\makecell{High}} & \textcolor{blue}{[R]} \textit{``The writer would not have been able to write the same or similar content without AI; relying solely on their own capability or other non-AI tools (such as a dictionary or traditional search engine) would not have led to comparable outcomes.''} \\
& & \textcolor{orange}{[W]} \textit{``You would not have been able to write the same or similar content without AI; relying solely on your own capability or other non-AI tools (such as a dictionary or traditional search engine) would not have led to comparable outcomes.''} \\
\midrule
\multirow{4}{*}{\rotatebox{90}{\makecell{\colorbox{Goldenrod}{\footnotesize{EFFORTFULNESS}}}}}& 
\multirow{2}{*}
{\makecell{Low}} & 
\textcolor{blue}{[R]} \rule{0pt}{3ex}
\textit{``The writer put in much less time and effort to complete the writing, thanks to AI.''} \\
& & \textcolor{orange}{[W]} \textit{``You put in much less time and effort to complete the writing, thanks to AI.''} \vspace{0.6em} \\
\cmidrule(lr){2-3}

& \multirow{2}{*}{\makecell{High}} & \textcolor{blue}{[R]} \rule{0pt}{4ex}\textit{``The writer put in a lot of time and effort to complete the writing, even with the help of AI.''} \\
& & \textcolor{orange}{[W]} \textit{``You put in a lot of time and effort to complete the writing, even with the help of AI.''} \vspace{0.6em} \\
\midrule
\pagebreak  
\multirow{8}{*}{\rotatebox{90}{\makecell{\colorbox{Goldenrod}{\footnotesize{INTENTIONALITY}}}}}
& \multirow{4}{*}{\makecell{Low}} & \textcolor{blue}{[R]} \textit{``The writer did not have a concrete blueprint of what they wanted the writing to look like, and they let AI take the lead and steer the direction of the writing.''} \\
& & \textcolor{orange}{[W]} \textit{``You did not have a concrete blueprint of what you wanted the writing to look like, and you let AI take the lead and steer the direction of the writing.''} \\
\cmidrule(lr){2-3}
& \multirow{4}{*}{\makecell{High}} & \textcolor{blue}{[R]} \textit{``The writer had a concrete blueprint of what they wanted the writing to look like, and they steered AI's generation toward that target.''} \\
& & \textcolor{orange}{[W]} \textit{``You had a concrete blueprint of what you wanted the writing to look like, and you steered AI's generation toward that target.''} \\
\midrule

\multirow{7}{*}{\makecell{\rotatebox{90}{\colorbox{Goldenrod}{\footnotesize{DIRECTNESS}}}}}
& \multirow{3}{*}{\makecell{Low}} & \textcolor{blue}{[R]} \textit{``The writer used AI's generation only as a reference, without directly incorporating it into their writing.''} \\
& & \textcolor{orange}{[W]} \textit{``You used AI's generation only as a reference, without directly incorporating it into your writing.''} \\
\cmidrule(lr){2-3}
& \multirow{3}{*}{\makecell{High}} & \textcolor{blue}{[R]} \textit{``The writer directly incorporated much of AI's generation as-is or with few changes into their writing.''} \\
& & \textcolor{orange}{[W]} \textit{``You directly incorporated much of AI's generation as-is or with few changes into your writing.''} \\

\end{longtable}

\section{Post-study questionnaire} ~\label{appendix:post-study-questionnaire}

Table~\ref{table:demographics} summarizes the participants' demographic information.

\begin{table}[htbp]
\small
\centering
\begin{tabular}{ >{\raggedright\arraybackslash}m{5cm} >{\raggedright\arraybackslash}m{5cm} }
\toprule
\textit{Questions} & \textit{Options [Number of participants]} \\
\hline
\textbf{What is your gender identity?}  & Female [368]  \newline
                        Male [345]  \newline
                        Other [11]  \newline
                        Prefer not to say [3]\\
\hline
\textbf{How old are you?}  & 18-24 years old [58] \newline
                    25-34 years old [198]   \newline
                    35-44 years old [172] \newline
                    45-54 years old [168]  \newline
                    55-64 years old [81] \newline
                    65+ years old [50] \\
\hline
\textbf{What is your ethnic background?}  & White [505] \newline
                     Asian [50] \newline
                      Native Hawaiian or Pacific Islander [2]
                      \newline
                      Hispanic or Latino [41]
                      \newline
                      Black or African American [100]
                      \newline
                      Native American [7]
                      \newline
                      Two or more [21]
                      \newline
                      Other [0]
                      \newline
                      Unknown [0]
                      \newline
                      Prefer not to say [1]\\
\hline
\textbf{What is your level of education?}  & Less than High School [3] \newline
                      High school [76]
                       \newline
                      Some college (no degree) [130]
                       \newline
                      Associate degree [74]
                       \newline
                      Bachelor’s degree [293]
                       \newline
                      Master’s degree [118]
                      \newline
                      Doctoral degree [20]
                      \newline
                      Professional degree (JD, MD) [13]
                     \newline
                      Prefer not to say [0]\\
\bottomrule
\end{tabular}
\caption{Demographic overview.}
 \Description{In this table, we describe the survey items we used in the post-study questionnaire for collecting participants' demographic information (Left). We also summarized the participants' demographics by groups (Right).}
\label{table:demographics}
\end{table}

Table~\ref{table:post-study-questionnaire} shows the survey items measuring self-efficacy of writing (adapted from~\citet{schwarzer1995generalized}), AI literacy (adapted from~\citet{wang2020factors}), internal judgment of AI use (adapted from~\citet{zhang2025secret}), and internal judgment of disclosure.

\begin{table}[htbp]
\small
\centering
\begin{tabular}{ >{\raggedright\arraybackslash}m{4cm} >{\raggedright\arraybackslash}m{9cm} }
\toprule
\textit{Covariates} & \textit{Questions} \\
\hline
\textbf{Self-efficacy of writing} \newline 5-point Likert scale \newline Not at all true---Exactly true  & When I am writing, I can always manage to solve difficult problems if I try hard enough.\smalladdlinespace \newline
                        When I am writing, I can remain calm when facing difficulties because I can rely on my coping skills.\smalladdlinespace \newline
                        If I am in a bind when writing, I can usually think of something to do.\\
\hline
\textbf{AI literacy} \newline 5-point Likert scale \newline Strongly disagree---Strongly agree & I can make use of AI to write. \smalladdlinespace \newline
                    I am confident in using AI.  \smalladdlinespace \newline
                    I can use a chatbot like ChatGPT to help me with writing. \smalladdlinespace \newline
                   I can craft a prompt specifically for a task I'm trying to perform.\\
\hline
\textbf{Internal judgment of AI use} \newline 5-point Likert scale \newline Strongly disagree---Strongly agree & I feel less competent when I need AI to assist me with writing. \smalladdlinespace \newline
                     I have doubts about the effectiveness of using AI in writing. \smalladdlinespace \newline
                      I think using AI for writing is immoral or unethical.\\
\hline
\textbf{Internal judgment of disclosure} \newline 5-point Likert scale \newline Strongly disagree---Strongly agree & I value transparency in communicating about my AI use. \smalladdlinespace \newline
                      I think that honesty about AI use is valuable.
                       \smalladdlinespace \newline
                      I think disclosing the use of AI is closely related to one's own value of transparency.
                      \\
\bottomrule
\end{tabular}
\caption{Survey items measuring self-efficacy of writing, AI literacy, internal judgment of AI use, and internal judgment of disclosure.}
\Description{In this table, we include all survey items that we used for measuring self-efficacy of writing, AI literacy, internal judgment of AI use, and internal judgment of disclosure in the post-study questionnaire.}
\label{table:post-study-questionnaire}
\end{table}

\section{Qualitative analysis and findings}
To supplement the logistic regression, we conduct a qualitative analysis on the participants' open-ended responses regarding why they consider disclosure to be (un)necessary. 

\subsection{Method} \label{appendix:method_qual}

Two of the authors familiarize themselves with the data and independently generate initial codes. Through ongoing discussions, they iteratively refine and expand the codes and co-construct a shared understanding of the data, and themes emerge during the process. Each response may be labeled with one or multiple themes to comprehensively capture the rationales. Our approach follows the reflexive thematic analysis method~\cite{braun2019reflecting}, which is commonly used in HCI research~\cite{he2025deservecredit, bowman2023using}. Note that the thematic analysis was conducted on all participants' responses without stratification by perspective, and therefore, the emerged themes capture insights from both readers' and writers' perspectives. 

\subsection{Findings}
\label{appendix:results_qual}
Four themes emerged in our thematic analysis. These themes supplement quantitative results (reported in Section~\ref{sec:results}) by providing richer contexts of the rationales and surfacing potential factors that were not part of our vignette study. Notably, Section~\ref{sec: qual_procedural} reaffirms our quantitative findings of procedural factors. 

\subsubsection{Purpose and scenario of reading and writing (N=129)} \label{sec: qual_purpose}
Participants note that disclosure may affect how readers use or interact with the written text, especially when readers have a certain purpose when reading.
For example, participants note that the reliability and trustworthiness of the writing content are particularly important when reading or writing will lead to direct actions (e.g., when one is reading to learn about knowledge that they will be tested on). For example, P13 (assigned scenario: educational article about American history) mentioned, \textit{``I think it's important for people to know where the information came from. Also, people might interact with it differently or feel less comfortable if they know AI was involved''}. Similarly, P65  (assigned scenario: computer software tutorial) mentioned, \textit{``Transparency helps maintain trust, especially if the tutorial influences how users interact with important software''}. Some participants apply heuristics, such as whether the written text is goal- or leisure-oriented, or whether it is commercialized, to gauge whether the writing scenario is ``formal'' or ``casual''. They then judge the necessity of disclosure based on the formality (according to their own standards) of the writing scenario. For example, P19 (assigned scenario: travel blog post) mentioned, \textit{``I don't think it's necessary to add that you used [AI] if it's just for light entertainment''}. P75 (assigned scenario: student’s personal statement) mentioned, \textit{``The admission officer should have the right to know if the writer knows grammar, spelling, and how to structure a sentence''}. 

\subsubsection{AI usage in writing (N=407)} \label{sec: qual_procedural}
Many participants directly mention one or more of the procedural factors that we manipulated in the vignettes (replaceability, effortfulness, intentionality, directness) as the reason that disclosure is (un)necessary. 
For example, P44 (assigned directness: low directness) explained why they think disclosure is unnecessary with \textit{``I only used the AI as a reference instead of having the AI do everything and write everything''}. 
People also gauge the necessity of AI disclosure according to their overall impression of the writer's AI usage, which is implied 
by the combination of procedural factors. For example, people judge the necessity based on their discretion of whether the writer can claim ownership of the work. 
P37 thought that disclosing their AI use is necessary because \textit{``I cannot try to pass off the work as my own with how much the AI assisted''}. 
P671 believed that disclosure is unnecessary after evaluating the extent of AI use and concluding that \textit{``[AI's] role was pretty minor''}.
Participants also consider whether AI's actual role in the writing is similar to (or different from) a non-AI tool or a human helper. For example, P595 thought that disclosure is unnecessary because \textit{``Using AI in this way isn't really any different than having a friend to bounce ideas off and give advice''}. 

\subsubsection{One's personal value and standard of disclosure (N=285)} \label{sec:qual_personal}
Some participants believe that there is no need to disclose as long as disclosure is not explicitly required or enforced. For example, P366 thought that disclosure is unnecessary simply because \textit{``There is no specific rule requiring disclosure. It does not appear to be against the rules to not disclose''}. Moral considerations play an important role in the perceived necessity of disclosure. For example, P381 said that \textit{``Anytime you use AI for anything, it should be admitted to and clearly states that AI was involved in doing it''}.  Several participants think that disclosure is unnecessary because they question the usefulness and benefit of disclosure. For example, P480 mentioned that \textit{``Disclosure does not add meaningful benefit to the audience''}. Similarly, P592 mentioned that \textit{``I don't think it's at all necessary to disclose my AI use because it doesn't matter to the reader whether I wrote the piece completely on my own or incorporated AI''}.

\subsubsection{Social factors (N=102)} \label{sec:qual_social}
Participants think about how the disclosed AI use might be perceived by others (i.e., social factors) when people judge the necessity of disclosure. They consider the bad consequences of (non-)disclosure. For example, P12 did not think disclosure is necessary because \textit{``The reader may be biased towards AI works''}. Conversely, some participants mentioned that disclosure is necessary because non-disclosure may lead to bad consequences if AI's involvement is found out by readers without the writer's proactive disclosure. For example, P128 thought that disclosure is necessary due to potential bad consequences of non-disclosure---\textit{``I think that it is important [to disclose], as if you do not, the reader can notice that you used AI, and so would undermine your credibility as a writer''}. On the other hand, people also take the good consequences of (non-)disclosure into consideration. For example, P402 thought that AI disclosure is necessary as \textit{``Being honest about how the content was created helps build trust''}. Sometimes people think that disclosure is unnecessary because they worry about readers' unacceptance of AI writing or writers who use AI. For example, P453 said that \textit{``Honestly, I would like to disclose my AI use. However, I feel like it could drive away potential readers who are critical of AI because they may assume that my story was more AI-assisted than it actually is, even if I'm honest about what I used it for''}. Some responses imply that the prevalence of AI use in writing rationalizes non-disclosure, probably because they think readers should assume that AI is used even when it is not explicitly disclosed. For example, P289 thought that disclosure is unnecessary because \textit{``It is a common practice''}. P386 similarly explained why they think disclosure is unnecessary with \textit{``We are in a very digital world where almost everyone has transitioned to using AI''}.

\paragraph{There are a few responses exhibiting a misunderstanding of the question we asked (N=7)}  For example, some participants talk about why they believe using AI in the writing scenario is (in)appropriate, which is not directly related to why disclosure is (un)necessary. We leave these responses out of our analysis.